\documentclass[useAMS,usenatbib]{mn2e}

\usepackage{lscape}
\usepackage{graphicx}
\usepackage{multirow}
\usepackage{amssymb}

\usepackage{multirow}

\title[The CR Transition.]{Observational indicators of the transition
  from fully convective stars to stars with radiative cores.}
\author[N.J.  Mayne.]{N.J. Mayne$^{1}$\thanks{E-mail:
    nathan@astro.ex.ac.uk
    (NJM)}\\
  $^{1}$ School of Physics, University of Exeter, Stocker Road,
  Exeter, EX4 4QL.
  \\
  \\
}
 
\begin{document}

\date{Accepted ?. Received ?; in
  original form ?}

\pagerange{\pageref{firstpage}--\pageref{lastpage}} \pubyear{2009}

\maketitle

\label{firstpage}

\begin{abstract}  
  We present a discussion of the similarities and key differences
  between the transition onto (at the turn-on) and away from (at the
  turn-off) the main sequence, the latter termed the Hertzsprung
  gap. Using a set of model isochrones and adopting an initial mass
  function leads us to predict a dearth of G-type stars for any star
  forming region. This is caused by the (relatively) constant spectral
  type at which the transition from a fully convective star to a star
  with a radiative core begins. We also present analysis of the
  details of this transition in the ONC. In particular we show that a
  gap in the photometric and spectral type distributions is centred
  on, and a change in the fractional X-ray luminosity and rotation
  rate distribution occurs approximately at, the position of a peak in
  radiative core size as a function of mass. Whilst photometric
  signatures of this transition are lost at ages over $\sim$20 Myrs,
  we show that changes in fractional X-ray luminosity and magnetic
  field configuration persist to older ages. Analysis of literature
  data show that the mass at which the change in fractional X-ray
  luminosity is observed decreases with age.
   \end{abstract}


\section{Introduction}
\label{intro}

The Hertzsprung gap (H gap hereafter) is a region of paucity of stars
separating the main-sequence (MS) and the post-MS red giant
phase. This gap is observed in colour-magnitude diagrams (CMDs) and HR
diagrams of galactic clusters \citep[see for
example][]{hoyle_1960}. The H gap is caused by a region of swift
evolution (instantaneous when compared to the MS lifetime of these
stars) in effective temperature ($T_{\rm eff}$) and therefore
colour. This swift change in temperature is, in turn caused by the
cessation of core hydrogen burning and the build up of helium ash
leading to a change from a radiative to a convective core. The
convective core then contracts, which as it is separated by an active
hydrogen burning shell, leads to the envelope expanding, and cooling.
The MS edge of this gap is located at the turn-off, and is well
predicted by nuclear burning models for MS and post-MS objects. Of
course, practically, the transition of stars through the gap is not
observed over time but actually as a region of swift $T_{\rm eff}$
evolution as a function of, for instance, mass or spectral type.

A similar gap has been observed for pre-MS stars
\citep{piskunov_1996,belikov_1997,stolte_2004,piskunov_2004,mayne_2007,bono_2010,cignoni_2010,rochau_2010}. In
this case the fully convective stars, contracting onto the MS,
eventually form a radiative core with the `ignition' of hydrogen
burning. Once the core is formed it expands relative to the total
stellar radius (which contracts overall after a brief expansion). The
transition, as with the H gap, leads to a swift evolution in $T_{\rm
  eff}$ (which again, as with the H gap, can be considered
instantaneous when compared to the MS lifetime and also the
contraction time). The MS edge of this transition is located at the
turn-on. Once more this is practically observed as a rapid change in
$T_{\rm eff}$ as a function of mass or spectral type.

The H gap and its pre-MS analogue are both caused by rapid evolution
in $T_{\rm eff}$, changes in the dominant energy transfer mechanism
within the core (either convective $\to$ radiative or vice versa) and
the cessation or initiation of core hydrogen burning. However several
key differences exist between these two transitions meaning they are
not simply the same process in reverse. Firstly, the location, within
a CMD is clearly different, with the turn-on and turn-off (if both are
present) separated in magnitude. Additionally, and most crucially for
this work, the stratification of the dominant energy transfer
mechanism within the stars differ across the H gap and its pre-MS
analogue. For the H gap, simplistically, the stars moves from the MS,
with an associated radiative core and convective envelope, to the
post-MS, with a convective core, active (radiative) hydrogen burning
shell and convective envelope. Essentially, the stars begins the
transition with one boundary or shearing layer and ends it with two,
with the final configuration complicated by the shell. For the pre-MS
transition the stars moves from a fully convective regime to the
development of a radiative core and associated, single, shearing or
tachocline layer.

The transitions in both cases will be a function of age, as the mass
of the star begining the transition decreases with time. This age
dependence, for the pre-MS transition, has been highlighted and
exploited in several studies. \cite{piskunov_1996} constructed
theoretical luminosity functions using convolution of pre-MS
isochrones with a Saltpeter IMF. This resulted in a dip (the gap)
bounded by two peaks, in the luminosity functions. The bounding peaks
were attributed to radius inversion in the birthline (on the pre-MS or
convective side, termed the R maximum) and stars reaching the MS
(termed H maximum). \cite{piskunov_2004} apply the work of
\cite{piskunov_1996} \citep[and the further work of][]{belikov_1997}
by using the so--called H feature (named thus as it represents the
`ignition' of hydrogen burning) to aid the ageing of several
clusters. \cite{cignoni_2010} describe a similar technique for
deriving cluster ages but by binning stars in magnitude. This is done
over increasing radii from the cluster core and identification of a
magnitude at which they find a peak followed by a fall in the number
of stars. The choice to bin the stars by magnitude, as opposed to
colour, means the method of \cite{cignoni_2010} and
\cite{piskunov_2004} will include an inherent degeneracy with
distance.

\cite{mayne_2007} identified the manifestation of the pre-MS H gap
within the CMDs of several young clusters. In this work the gap,
termed the radiative convective (RC) gap, is identified and its causes
in terms of the change in dominant core energy transfer mechanism
explained. Furthermore, \cite{mayne_2007} suggest that the gap should
be measured in colour, as its size is a function of age, meaning it
would be free of distance degeneracy problems. The pre-MS H gap has
also been used by \cite{bono_2010} and \cite{stolte_2004} to aid
isochrone fitting. \cite{bono_2010} define features of the isochrone
called the main-sequence turn-off (MSTO) and main-sequence knee
(MSK). In contrast, \cite{stolte_2004} use the shape of the so--called
MS/pre-MS transition region.

As a cluster ages the photometric signature of the RC or pre-MS H gap
is lost. This is as the turn-on becomes less well separated from the
pre-MS with age, and the transition occurs for lower mass stars
resulting in a less signifcant and slower change. Practically, this
means that any photometric signature is lost above an age of around 20
Myrs. However, as the pre-MS stars move from the fully convective
regime to one with a radiative core, several secondary characteristics
are changed as a consequence. As shown in the work of
\cite{endal_1981} once a radiative core forms this will decouple from
the convective envelope. Therefore, as these stars undergo a rapid
transition one might expect a signature in the rotational period
distributions across the transition. Additionally, the decoupling of
the core will lead to a change in the magnetic field generation within
the star, from fully convective to a tachocline driven mechanism, as
the decoupling creates a shearing layer. This in turn should lead to
signatures in the X-ray and variability signatures of the stars across
the transition \citep[see][for discussion of the
latter]{saunders_2009}.

The $T_{\rm eff}$, magnetic field generation mechanism, angular
momentum distribution and X-ray emission are all changed due to the
transition from fully convective star to one with a radiative
core. Therefore, throughout this work we term the pre-MS H gap
analogue the CR transition to highlight the importance of the change
in dominant energy transfer mechanism. In this paper we explore the
details of the CR transition with particular reference to the growth
of the radiative core, using theoretical data and data for stars in
the ONC. The models used throughout this paper are those of
\cite{siess_2000}. In Section \ref{theory} we illustrate the
photometric signatures of the CR transition and highlight implications
for spectral type distributions. In Section \ref{onc} we present
signatures of the CR transition in data of the ONC and show that these
match predictions of the growth of a radiative core. In Section
\ref{other} we present a summary of some literature results containing
indications of the CR transition in several clusters and in several
observables. Finally we conclude in Section \ref{conclusions}.

\section{Theory of the pre-MS H gap, or CR transition}
\label{theory}

\begin{figure}
\includegraphics[scale=0.32,angle=90]{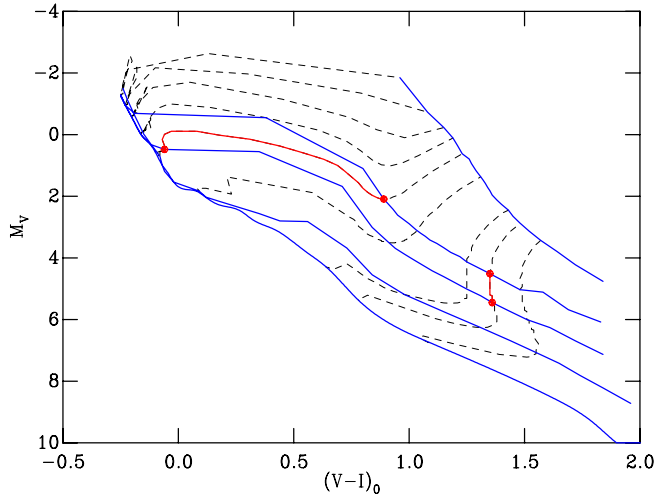}
\caption{Figure showing the isochrones of \citet{siess_2000} at 1, 3,
  and 13 Myrs, with the static-MS and 0.1 Myrs as bounding lines, as
  solid blue lines. The mass tracks of \citet{siess_2000} for 7, 6, 5,
  4, 3, 2, 1.2, 1.0, and 0.8 $\rm M_{\odot}$ are shown as dashed black
  lines. The evolution between 1 and 3 Myrs of 3 and 1 $\rm M_{\odot}$
  stars are shown as filled circles, with the track highlighted in
  red. \label{theory_ps}}
\end{figure}

The key characterisitic of the CR transition is its age dependence,
with several authors using the location of the gap (or associated
features) to derive ages for clusters
\citep{piskunov_2004,mayne_2007,cignoni_2010,bono_2010} The age
dependency of the size is a result of the convective part of the
pre-MS isochrones moving closer to the MS with age. Additionally, the
mass at which the transition is occuring reduces leading to a slower
and less tumultuous transition. Figure \ref{theory_ps} is a
reproduction of Figure 19 from \cite{mayne_2007}. This shows that the
separation between the head of the pre-MS and the tail of the MS
decreases with age. The isochrones of 1, 3 and 13 Myrs and the
early-MS (0.1 Myrs is shown as a pseudo-birth line) from
\cite{siess_2000}, are shown as well as the mass tracks for 7, 6, 5,
4, 3, 2, 1.2, 1.0, and 0.8 $\rm M_{\odot}$. The early-MS isochrone
differs from the zero-age-main sequence (ZAMS) isochrone for higher
mass stars as these stars undergo a cycle of CN burning which involves
evolution above, in terms of a CMD, the ZAMS until this cycle reaches
equilibrium. The early-MS simply shows the position of stars after
they have `settled' onto the, relatively, temporally static-MS. The
evolution between the 1 and 3 Myr isochrones of a 1 and 3 $\rm
M_{\odot}$ star is highlighted, in Figure \ref{theory_ps}, by the
solid (red) tracks and circles. By comparing the evolution of the two
highlighted masses one can see that in a relatively short period of
time (2 Myr) the more massive star has moved a significant distance
across the CMD and joined the MS, whilst the lower mass objects has
simply dropped slightly in magnitude. Additionally, the size in colour
of this transition region is a function of the colours and magnitudes
used. For instance the separation between pre-MS and MS is larger in a
$M_J, (J-K)_0$ CMD, meaning Infrared (IR) studies of the CR transition
region could yield easier detectability. Indeed, the pre-MS/MS
transition region was highlighted in an IR CMD by \cite{stolte_2004}
as a visual aid to fitting isochrones to IR photometry of NGC 3603.

\begin{figure}
\includegraphics[scale=0.32,angle=90]{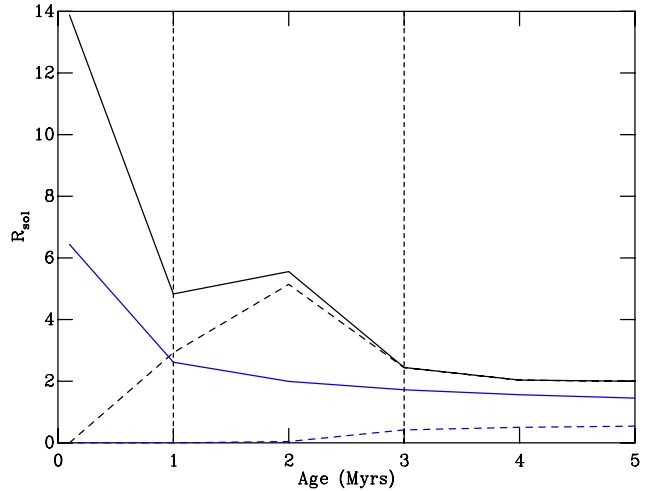}
\caption{Figure showing the stellar radii (solid line) and radius of
  the radiative core (dashed line) of the 1 and 3 $\rm M_{\odot}$
  (black and blue respectively) models of
  \citet{siess_2000}. \label{gap_age}}
\end{figure}

Figure \ref{gap_age} shows the stellar radius and radius to the bottom
of the convective envelope as a function of age for the 1 and 3 $\rm
M_{\odot}$ models of \citep{siess_2000}. The vertical dotted lines in
Figure \ref{gap_age} highlight the ages of 1 and 3 Myrs. As the star
moves across the CR transition, between 1 and 3 Myrs for the 3 $\rm
M_{\odot}$ model, the core grows within the star and the total stellar
radii undergoes a temporary expansion, but contracts overall. Once the
radiative core grows to almost the size of the star the transition is
complete and the star joins the MS. Across the same age range the 1
$\rm M_{\odot}$ model simply contracts slightly as it follows a
Hayashi track. Of course, as discussed, practically we do not observe
the transition as a function of time, but rather as a snapshot with
the transition apparent as a function of mass or spectral type.

\begin{figure}
\includegraphics[scale=0.32,angle=90]{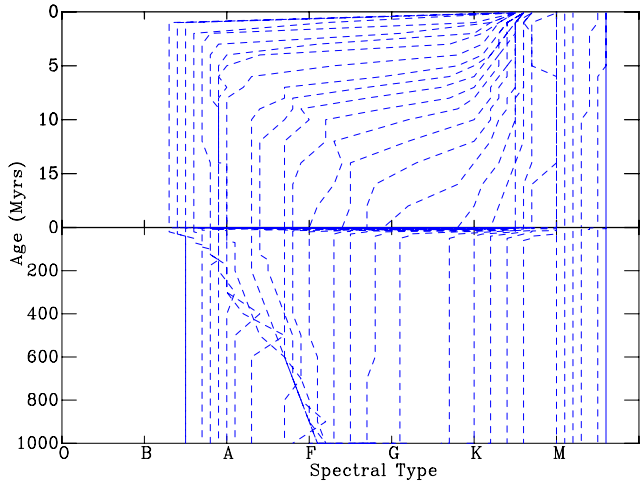}
 \caption{Figure showing spectral type as a function of stellar age
   for the range of masses available from the isochrones of
   \citet{siess_2000}. The top panel shows ages from 0 to 20 Myrs and
   the bottom panel up to 1000 Myrs.\label{spec_type_evol}}
\end{figure}

Whilst the CR transition can be observed in a CMD deriving masses,
using isochrones, for individual stars across the region is
unreliable. As the gap is caused by a rapid evolution in $T_{\rm eff}$
a signature will also be apparent in the spectral types of the
stars. Figure \ref{spec_type_evol} shows the spectral types of stars
as a function of age for each mass available from the models of
\citet{siess_2000}. The top and bottom panels simply show different
age limits to highlight the changes for different stages of stellar
evolution.

The top panel of Figure \ref{spec_type_evol} shows the size of the CR
transition region, areas where the tracks tend to horizontal,
decreasing as a function of age. Interestingly, the fully convective
edge of the transition stays at around late-G, whilst the MS edge
moves from A to G. Furthermore, the lower panel shows that even for
the much older sequences the tracks are well seperated across the G
spectral class. This is not caused by a resolution issue as the mass
increments in the models are constant across this range. This
essentially means that if we assume a smooth, flat,
Initial-Mass-Function (IMF) there should be a detectable paucity of
G-type stars within a spectral type histogram of any young cluster or
star forming region. Practically, one might expect this paucity may be
smoothed out when convolved with a reasonable IMF, however it is still
apparent in the ONC, as shown in Section \ref{onc} and Figure
\ref{onc_spec_core}.

\section{Gap in the ONC}
\label{onc}

In this section we explore the connection between the signatures of
the CR transition for the ONC and the predicted growth of the
radiative core.

\subsection{Photometry and Spectral Types}
\label{phot_spec}

\begin{figure}
\includegraphics[scale=0.32,angle=90]{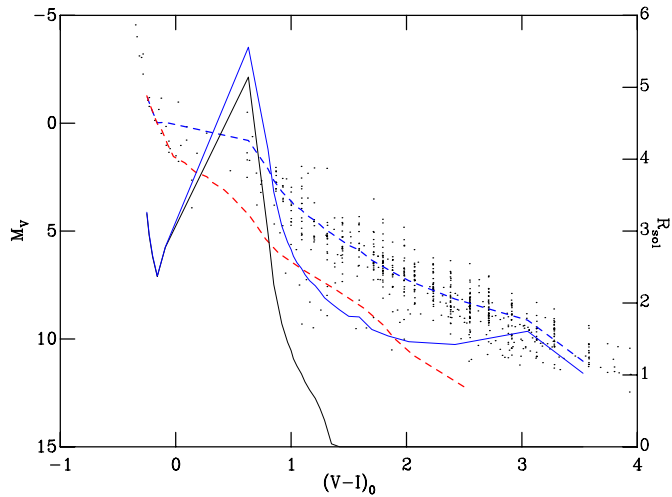}
\caption{Figure showing spectrally dereddened optical photometry in a
  $M_V$ (left Y-axis), $(V-I)_0$ CMD using data from
  \citet{hillenbrand_1997} and the 2 Myr and static-MS isochrones from
  \citet{siess_2000}. The solid black and blue lines show the absolute
  size of the radiative core (or radius to the convective envelope,)
  and radii of the stars, respectively (as the right Y-axis), as
  predicted by \citet{siess_2000}, for an age of 2
  Myrs.\label{onc_phot_core}}
\end{figure}

The ONC \citep[$\approx$2 Myrs, $\approx$400 pc][]{mayne_2008} is one
of the clusters where the RC gap was observed
\citep{mayne_2007}. Absolute photometry of the ONC is shown in Figure
\ref{onc_phot_core}, where a distance modulus ($\rm dm$) of 7.96 has
been adopted \citep{mayne_2008}. The optical photometry and
spectroscopy of the stars with membership probabilities in excess of
80$\%$ from \cite{hillenbrand_1997} have been used to construct an
$M_V$ (left Y-axis), $(V-I)_0$ CMD, Figure \ref{onc_phot_core}. For
the optical data each star was individually corrected for extinction
(in cases where an individual extinction was not available a mean
extinction was applied). The 2 Myr and the static-MS of
\cite{siess_2000} (dashed lines) are also shown. Additionally, the
radius to the bottom of the convective envelope, or the size of the
radiative core, and the predicted radii of the stars are shown (solid
lines, right Y-axis), from the models of \cite{siess_2000}, for an age
of 2 Myrs.

Figure \ref{onc_phot_core} shows a clear sparsity of stars in the region
separating the populous, photometrically spread, pre-MS members and
the fewer (due to the IMF predicting a reduction in the number of
stars as mass increases) less photometrically scattered MS members,
spanning a region from around $(V-I)_0\approx$ 0.0 to 0.90. The solid
black line shows that the stars become more dispersed at the first
appearance of the radiative core. In this figure we observe the growth
of the core as a function of mass or spectral type, not over
time. However, clearly, as the mass increases (towards brighter
magnitudes) the core (and the star) size goes through a
`spike'. Essentially, the radius of the stars, and relative size of
their cores, within the gap goes through a region of rapid increase as
a function of mass. At higher masses still the core encompasses almost
all of the star, and the stars have settled onto the MS, with a small
contraction as a function of mass. As the ONC lies in front a dense
molecular crowd there is little contamination from background
members. Additionally, this sample, it is reasonable to assume, is
photometrically complete down to around $M_V\sim 9$ (i.e. 2 mags above
$M_{\rm V, \emph{lim}}$), which is well below the noted
gap. Saturation may cause the loss of some brighter objects but this
does not effect our results, in fact the addition of more brighter
stars would strengthen our conclusions.

\begin{figure*}
\includegraphics[scale=0.6,angle=90]{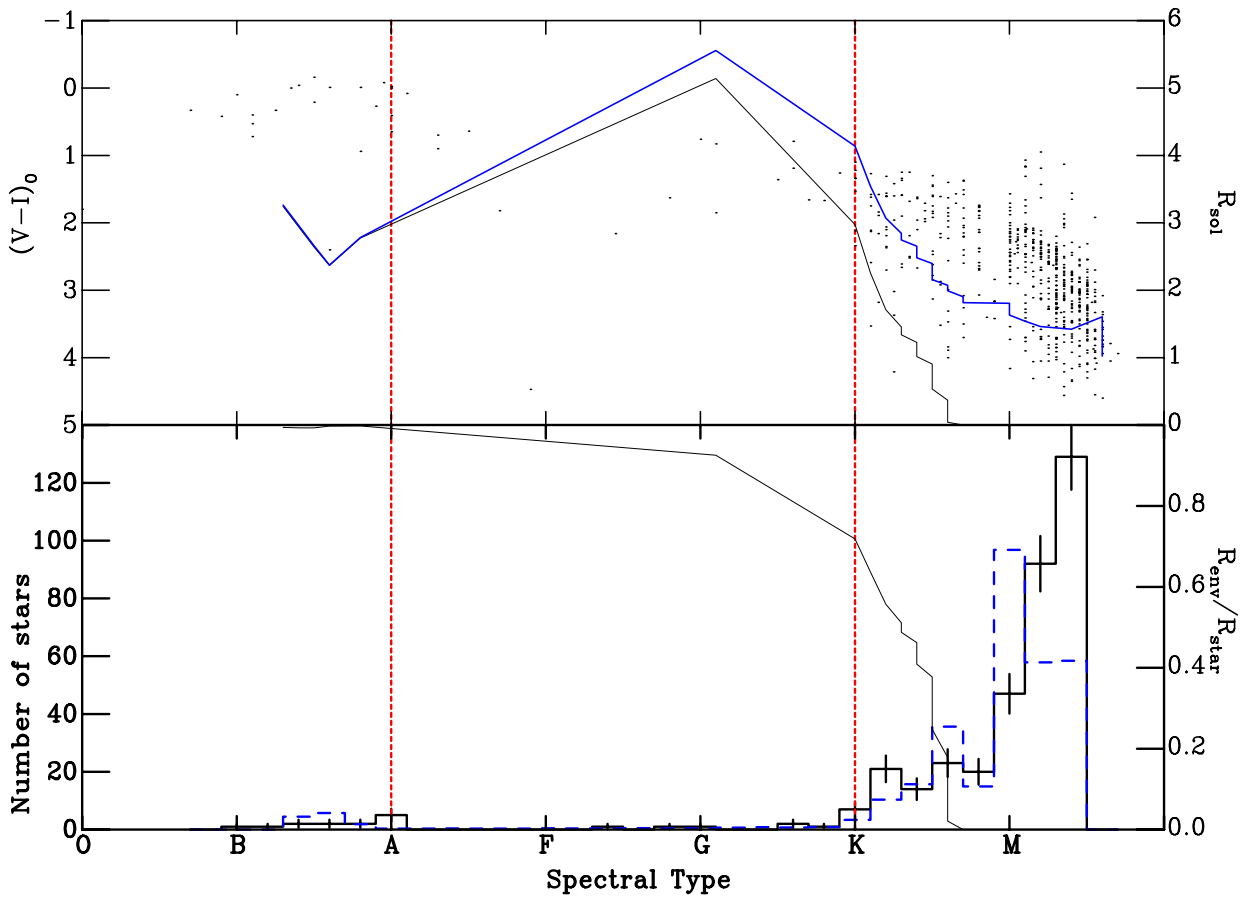} 
\caption{Figure showing the spectral types of stars in the ONC from
  \citet{hillenbrand_1997}. The top panel shows the spectral types
  (dots) against colour (left Y-axis) and the bottom panel shows the
  number of each spectral type (left Y-axis, solid histogram, vertical
  bars are uncertainties). The solid black lines in the top and bottom
  panels then show the absolute ($R_{\odot}$) and relative sizes of
  the radiative core from the models of \citet{siess_2000},
  respectively, for an age of 2 Myrs. The solid blue line in the top
  panel shows the predicted radii of the stars \citep{siess_2000},
  again at 2 Myrs. The dashed blue histogram in the lower panel shows
  the simualted spectral type distribution using isochrones of
  \citet{siess_2000} in conjunction with a Kroupa IMF. Finally, the
  vertical (red) dotted lines show the approximate positions of the
  start and end of the gap in spectral types, for illustrative
  purposes only.\label{onc_spec_core}}
\end{figure*}

The data of \cite{hillenbrand_1997} also includes spectral types
derived from high resolution spectra (not from conversions of
photometry). Figure \ref{onc_spec_core} shows the histogram of the
number of stars classified as each spectral type by
\cite{hillenbrand_1997} (left Y-axis, vertical bars are uncertainties)
in the bottom panel (solid line) alongside the predicted spectral type
distribution derived from a Kroupa-IMF and the models of
\cite{siess_2000} (dashed line). For stars with several or large
ranges of determined spectral types a median spectral type was
chosen. The top panel of Figure \ref{onc_spec_core} simply shows the
spectral type and intrinsic colour of each star (left
Y-axis). Additionally, the top and bottom panels show the predicted
\citep{siess_2000} absolute and relative sizes of the radiative core
(right Y-axis), respectively (solid black lines), for an age of 2
Myrs. The top panel also shows the predicted radii of the stars
\citep[solid blue line,][]{siess_2000}, again at an age of 2
Myrs. Finally, the approximate edges of the paucity of spectral types
are also shown (vertical dotted red lines). The predicted spectral
type distribution has been derived by interpolating a Kroupa-IMF for
spectral type using the 2 Myr isochrone of \cite{siess_2000}, and
adopting a total visible system mass of around 380 $\rm M_{\odot}$ in
the range 0.1 $\rm M_{\odot}< \rm M_* <$7.0 $\rm M_{\odot}$. This is a
reasonable total mass estimate given that the catalogue contains
around 600 members and the total number of members is around six or
seven times this, $\geq$4000 giving an estimated total system mass
$\sim$2500 $\rm M_{\odot}$, consistent with the lower limit of
$\sim$2000 $\rm M_{\odot}$ \citep{olczak_2006}.

The top and bottom panels of Figure \ref{onc_spec_core} shows a clear
sparsity of the observed spectral types between approximately K0 to
B8. We have applied the Hartigan dip test to the histogram data from
the lower panel of Figure \ref{onc_spec_core}
\citep{hartigan_1985}. This results, after extrapolation from the
results tables of \cite{hartigan_1985}, in a probability of less than
$<5$\% \footnote{This value is taken from the table for 200
  datapoints, whereas we have 625, increasing the number of datapoints
  results in a lower percentage for a given KS statistic
  value. Therefore in reality the result is even stronger.} that the
distribution was drawn from a uni-modal underlying distribution. This
effectively shows that a dip is present. Additionally, we have tested
the the statistical significance of the observed (solid histogram) and
simulated (dashed blue histogram) spectral type distributions, in
their cumulative forms, being drawn from different populations. A
simple $\chi^2$ test \footnote{A 2--sided KS test is unsuited to these
  data as spectral types are discrete and the KS test models
  continuous distributions} results in the rejection of the null
hypothesis that the two, cumulative, distributions where not drawn
from the same underlying population at around 80\%.

More importantly, the growth of the radiative core within the star in
both relative, (bottom) and absolute (top) size, appears to coincide
with the beginning of the dip or gap. The initial appearance of the
core ($\sim$K6--7) coincides with a reduction in the density of the
stars, as shown in the top panel of Figure \ref{onc_spec_core} (and
apparent in Figure \ref{onc_phot_core}). Furthermore, the bottom panel
shows that as the core grows past around 70\% of the total stellar
radii the number of stars at those, and earlier, spectral types
($\sim$K0) decreases dramatically, illustrated by the rightmost
vertical (red) dotted line. Additionally, a spike or peak in the size
of the core (in part caused by expansion of the star itself as a
function of spectral type) just precedes the region of lowest density
of stars ($\sim$G).

In summary, the first appearance (in mass or spectral type) of a
radiative core creates a fall in the density of stars in both an
example CMD (Figure \ref{onc_phot_core}) and spectral type
distributions (Figure \ref{onc_spec_core}). As one moves towards
larger masses (or earlier spectral types) the gap becomes much
clearer, and the density drop more significant, once the core composes
around 70\% of the entire star. Additionally, as the radiative (and
total stellar radius) goes through a peak in its size (as a function
of mass or spectral type), the density of stars appears to be at its,
approximate, lowest. Finally, once the core encompases almost all of
the stellar interior the density of stars increases slightly as they
settle onto the MS.

\subsection{X-ray Luminosity and Angular Momentum}
\label{x_ray_ang}

Clearly, the growth of the radiative core coincides with significant
changes in the surface temperature and therefore spectral type of the
star. The reverse process, in the post-MS H gap, has as similar
effect. Crucially, however the structure within the stars across these
transitions in markedly different. The transition from MS star to
post-MS giant includes the generation of, effectively, an extra
shearing or tachocline layer as the convective core is seperated from
the envelope by an active shell. For the CR transition the star moves
from a completely convective regime to the generation of a boundary or
shearing layer. Therefore, in the pre-MS case one would expect more
significant changes in stars magnetic field, and therefore X-ray
emission.

\begin{figure*}
\includegraphics[scale=0.6,angle=90]{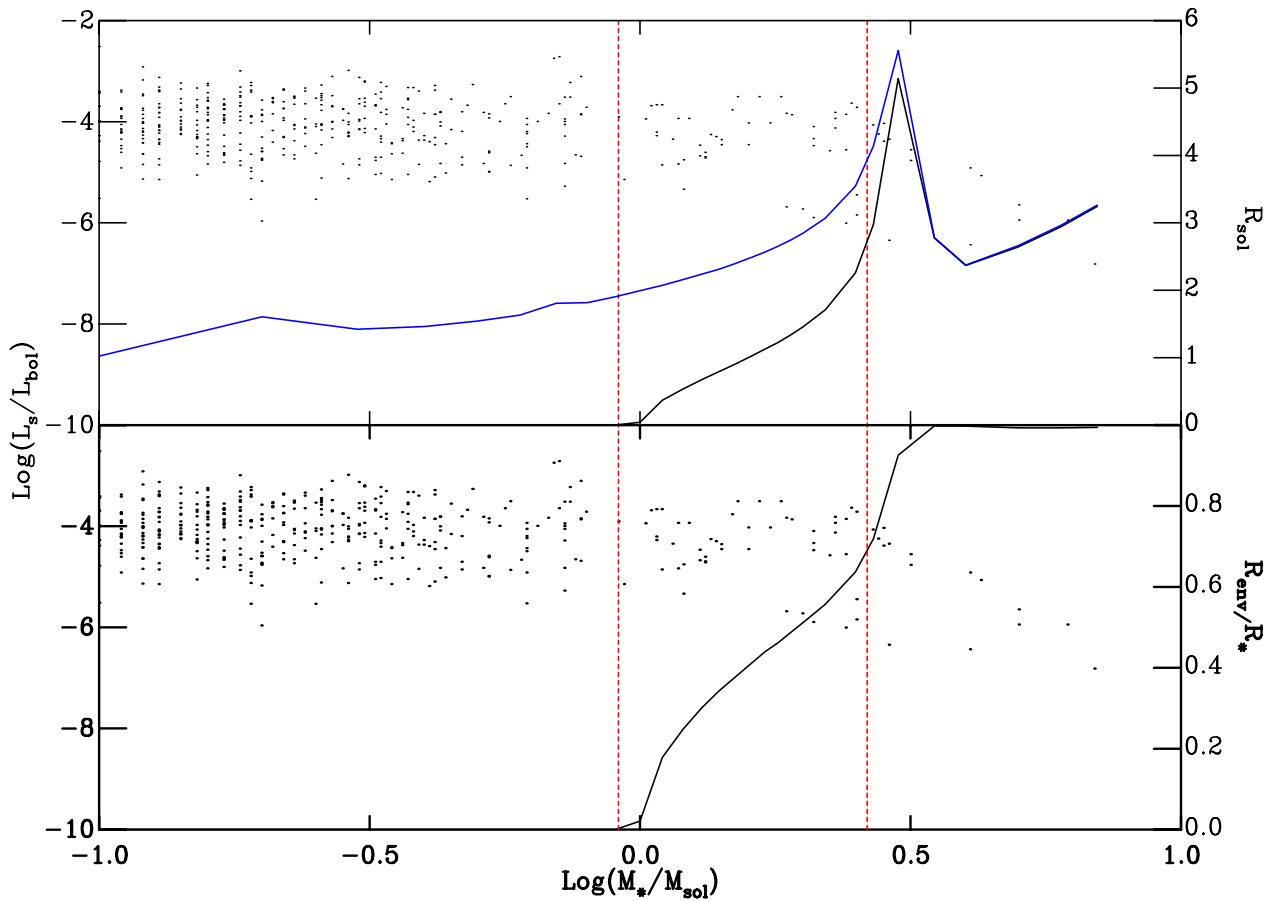} 
\caption{Figure showing (log) stellar mass against the (log) X-ray
  luminosity relative to the bolometric luminosity (left Y-axis) for
  the data of \citet{feigelson_2002}. The top and bottom panels then
  show, as solid black lines, the absolute and relative sizes of the
  radiative core from the models of \citet{siess_2000} respectively,
  for an age of 2 Myrs. Additionally, the top panel shows the
  predicted radii of the stars \citep{siess_2000} as a solid blue
  line, again for an age of 2 Myrs. Finally, the vertical (red) dotted
  lines, from left to right, mark the (log) masses at which the core
  first appears and encompasses 70\% of the
  star.\label{onc_xray_core}}
\end{figure*}

Figure \ref{onc_xray_core} shows the relative (to the bolometric
luminosity) X-ray luminosity (left Y-axis) against stellar mass for
stars in the ONC from the data of \cite{feigelson_2002}. The top and
bottom panels also show the absolute and relative sizes of the
radiative core (right Y-axis), respectively, from \citet{siess_2000},
at an age of 2 Myrs. Additionally, the top panel shows the predicted
radii of the stars \citep{siess_2000}, again at an age of 2 Myrs
(solid blue line). Finally, (from left to right) the (log) masses at
which the core first appears and encompasses around 70\% of the star
are indicated (vertical red dotted lines).

Figure \ref{onc_xray_core} shows that as the mass of the stars in the
ONC increases the relative X-ray luminosity remains approximately
constant, even when a core first appears at $\sim0.9 \rm M_{\odot}$
(leftmost vertical, red, dotted line), until around $\sim2.5-3$ solar
mass. Once the core encompasses around 70\% of the star, illustrated
by the rightmost vertical red dotted line, at a mass of $\sim2.6 \rm
M_{\odot}$ ($\sim$K0--K1), there is a sharp deline in the fractional
X-ray luminosity. This occurs at masses just preceding the `spike' in
core (and star) radial size. This suggests that the X-ray emission,
and therefore magnetic field topology is changing within the star as
the core develops. This is expected as the development of a core
creates a shearing or tachocline layer, leading to a transition from a
fully convective to a Sun--like dynamo. Indeed, \cite{feigelson_2002}
discuss the data shown in Figure \ref{onc_xray_core} in terms of the
relative strengths of radiative and fully convective dynamos. As X-ray
activity is used a selection tool for pre-MS populations this result
is unsurprising. This leads us to predict that Zeeman Doppler imaging
across the CR transition will lead to a marked change in the surface
magnetic field structure. Interestingly, the switch in magnetic dynamo
does not appear to occur until the radiative core comprises most of
the stellar interior.

As core--envelope decoupling is a natural consequence of the
generation of a radiative core one would also expect a change in
angular momentum distribution with the onset of radiative core
development. Evidence for a change in the rotation velocities across
this CR transition is shown and discussed in \cite{mcnamara_1990}.
\cite{mcnamara_1990} (albeit with a limited sample) discovers a change
in the $v\sin i$ distributions for stars classified as fully
convective or radiative with $\rm M_*>$1.5 $\rm M_{\odot}$ or $\rm
M_*<$1.5 $\rm M_{\odot}$. The distribution of rotation velocity was
shown to change across the CR transition with stars having radiative
cores rotating, overall, faster than their fully convective counter
parts. Figure 4 of \cite{mcnamara_1990}, shows distributions with
modal values of $<$25 kms$^{-1}$ and no stars rotating faster than 85
kms$^{-1}$ for the fully convective stars and a flatter much broader
distribution out to 150 kms$^{-1}$ for the higher mass stars with
radiative cores. In addition \cite{mcnamara_1990} also shows, with a
larger sample of stars a similar effect for Orion Ic. Again, this
suggests that further studies of the rotation rate distributions
either side of the CR transition will yield clear differences between
the populations.

\section{Other clusters}
\label{other}

A clear RC gap has also been found in the CMDs of several other
clusters such as NGC 2264, h and $\chi$ Persei and NGC 3603
\citep{mayne_2007,stolte_2004,rochau_2010}.

A further probe of the CR transition through analysis of variability
is presented in \cite{saunders_2009}. \cite{saunders_2009} show that
the fraction of stars with significant photometric variability reduces
as one crosses the RC gap in h Persei. Photometric variability in
pre-MS stars is generally caused by long lived surface cool spots
\citep{herbst_2007}. As these surface spots require large scale and
long lived surface magnetic fields their dissapearance is indicative
of a change in magnetic field topology. Essentially, the transition in
magnetic dynamo leads to a change from large scale ordered fields to
smaller scale fields \citep[see discussion and references
in][]{saunders_2009}. Again, this is unsurpising given that
significant photometric variability is often used as a pre-MS
selection tool.

As we have shown in Section \ref{onc} a decline in the relative X-ray
luminosity coincides with the growth of a radiative core, for the
ONC. Evidence for a similar decline can be found in the literature for
several other clusters. Moreover, if one compares the mass at which
the transition occurs there is tantalising evidence of the age
dependence of the CR transition. Figure 11 of \cite{dahm_2007} shows a
clear change in the $L_x/L_{bol}$ relationship, of stars in NGC2264
\citep[$\approx$3 Myrs,][]{mayne_2007}, at around a mass of 2 to 2.5
$\rm M_{\odot}$ or spectral type of K2--G7. \cite{currie_2009} find a
clear transition in the ratio of soft X-ray luminosity to the stellar
bolometric luminosity ($L_x/L_{bol}$) distribution at $(V-I)_0$ of
around 0.8, for stars in h and $\chi$ Per \citep[$\approx$14
Myrs]{mayne_2007,currie_2009}. As is shown in Figure 5 of
\cite{currie_2009}, this is around a spectral type of K0 or mass of
$\sim$1.3 $\rm M_{\odot}$ \citep{siess_2000}. A similar change in
gradient is evident in the fractional X-ray luminosity curve of
NGC2537 at $\approx$40 Myrs \citep{oliveira_2003}. The lower panel of
Figure 7 of \cite{jeffries_2006} shows that this change occurs at
$V-I$$\approx$0.75, for an age of 40 Myrs with minimal extinction
\citep[$E(B-V)=$0.04]{mayne_2008} this equates to a spectral type of
K0 and mass of 1 $\rm M_{\odot}$ \citep{siess_2000}. The transition is
also observable in the Pleiades Figure 14 from \cite{stauffer_2003}
shows a clear change in gradient of the fractional X-ray luminosity
against colour (or spectral type) within this cluster. The fractional
X-ray luminosity rises steeply from a $B-V\approx$0.5 to a peak at
around $B-V\approx$ 0.9, which given the small reddening to the
Pleiades \citep[$E(B-V)\approx$0.04][]{stauffer_2003}, equates to a
similar $(B-V)_0$. The places the CR transition at a spectral type of
around K2 or 0.9 $\rm M_{\odot}$. Finally, NGC2516 at $\approx$140
Myrs shows a softer transition in fractional X-ray luminosity. Figure
8 of \cite{pillitteri_2006} shows the change at a $(V-I)_0
\approx$1.00, at an age of 140 Myrs this is around a spectral type of
K4 and mass of 0.8 $\rm M_{\odot}$ \citep{siess_2000}. Therefore, we
can see that the mass at which the transition occurs decreases with
age, as expected from the models. The transition occurs at
approximately $2-3$ ($\sim2.6$), $2-2.5$, $1.3$, $1$, $0.9$ and $0.8$
$\rm M_{\odot}$ for the ONC ($\sim2$ Myrs), NGC2264 ($\sim3$ Myrs), h
and $\chi$ Per ($\sim14$ Myrs), NGC2537 ($\sim40$ Myrs), the Pleiades
($\sim125$ Myrs) and NGC2516 ($\sim140$ Myrs), respectively. These
masses, in all cases, are well matched by the masses at which the
growth of the radiative core undergoes the peak as shown for the ONC
in Figure \ref{onc_xray_core}.

\subsection{Magnetic Field Measurements}
\label{mag_field}

The fully convective dynamo, although poorly understood, produces
magnetic fields with different characteristics to the relatively well
understood solar-type dynamo which is based on a shearing layer
between radiative core and convective envelope. Recent studies using
Zeeman Doppler imaging have revealed a striking transition in the
stellar surface magnetic field across the CR transition for a
population of field stars \citep{donati_2008,morin_2008}. This
effectively means that indicators for the CR transition can be
detected even for the field stars which are many Gyrs old.

\cite{donati_2008} and \cite{morin_2008} observed Stokes \textit{V}
and \textit{I} line profiles of stars either side of the CR transition
region. They showed that as one moves across the CR transition the
stellar surface magnetic field undergoes a sharp transition. For fully
convective stars the field is multipolar \citep[not
dipolar,][]{gregory_2009} and retains large--scale structure
\citep{dobler_2006}. This enables long--term support of magnetically
controlled cool spots covering larger areas of the stellar surface
resulting in significant photometric variability \citep[see][for an
extended discussion of these results]{saunders_2009}. For those stars
with radiative cores the surface magnetic field is dominated by small
scale structures. This transition is observed to occur between masses
of around 0.35 to 0.4 $\rm M_{\odot}$, at or just after the lowest
mass at which a radiative core is predicted to have begun forming. It
is perhaps surprising that the switch in magnetic field is coincident
with the first appearance, as one increases in stellar mass, of a
radiative core. This suggests that even for a radiative core with a
small radial extent, relative to the star, the magnetic field
generation is dominated by its interaction with the envelope, and not
a convective dynamo within the envelope. This may well be explained by
the age of these stars. Essentially, the radiative core of these stars
has been present for many Gyrs meaning the dynamo based on the
shearing layer can slowly overpower the convective dynamo. The
efficiency of the fully convective dynamo mechanism has been shown to
be strongly dependent on rotation rate
\citep{dobler_2006}. Conversely, solar type dynamos, based on a
shearing layer and differential rotation, have been shown to be only
weakly dependent on the surface rotation rate \citep{kuker_2005}. It
has been known for some time that pre-MS stars spin up as they
approach the MS, with a peak in rotation rate of around 50 Myrs, and
then spin down as they evolve onto, and along the MS
\citep{endal_1981}. Therefore, for a field population of many Gyrs one
would expect the convective component of the magentic dynamo to be
minimised and the surface magnetic field generation to be dominated by
a solar type dynamo.

\section{Conclusions}
\label{conclusions}

In this work we have described the similarities and crucial
differences between H gap and its pre-MS analogue. We have used the
models of \cite{siess_2000} to explain the appearance of a gap in the
CMDs of young clusters \citep[previously termed the RC
gap][]{mayne_2007}, of age less than around 20 Myrs. Furthermore, we
have shown that the models of \cite{siess_2000} convolved with a
Kroupa-IMF lead to a prediction of a dearth of G-type stars in star
forming regions. This is as the convective side of the transition from
a fully convective star to the generation of a radiative core is at a
roughly constant spectral type as a function of age.

We have termed the pre-MS analogue of the H gap the CR transition. We
have then explored signatures of this transition in data of the ONC
and their relation to the growth of a radiative core. This resulted in
a clear change in not only in $T_{\rm eff}$ (and therefore colour and
spectral type) but also in fractional X-ray luminosities at a location
where the core growth is predicted to peak \citep{siess_2000}. We have
also highlighted literature evidence of changes in the angular
momentum distribution of stars across this CR transition in the ONC,
as would be expected from the inevitable core--envelope
decoupling. Furthermore, we have collected and repeated evidence in
the literature of transitions in the photometry of other clusters and
the variability of h and $\chi$ Per, with variability dying out across
the gap. Moreover, evidence from the literature of changes in the
fractional X-ray luminosity for several clusters reveals an age
sequence, with each transition well matched by predictions of the core
growth.

The clear extension is to use this exquisitely model dependent piece
of stellar evolutionary physics to fine--tune the models and constrain
associated predictions. This of course will both require, and help
develop a better model of convection in young (and old) stars by
providing an excellent boundary condition with respect to
thermodynamic and magnetic behaviour. Precise and multi-wavelength
observations across this CR transition region for a range of stellar
populations will lead to concrete constraints and boundary conditions
of great utility to models of convection and magnetic field generation
in stars. In addition, Zeeman Doppler imaging across the CR transition
for young stars is predicted to show a sharp transition in the
structure of the surface magnetic field. Further study of this aspect
may prove critical in the development of theories of magnetic field
generation in stars. We also predict that across the CR transition
region rotational velocity distributions should show a change, with
the radiative objects, generally, rotating faster than their fully
convective counterparts.

The CR transition region has been shown to be easily observable and is
delicately dependent on the stellar model. This transition demarcates
the boundary between the relatively well established nuclear burning
regime and the poorly understood convective regime. Therefore,
extended study of this phenomenon and its subsequent use as a
theoretical constraint should yield conclusive results with respect to
the thermodynamic, angular momentum, and magnetic field generation
evolution in young and old stars.

Finally, as shown in \cite{mohanty_2008} interaction between a young
star and circumstellar disc depends pivotally on the topology of the
stellar magnetic field. Accretion and disc interaction appears to be
maximised for stars with large scale ordered multi-polar fields
\citep[see discussion in][]{mohanty_2008}. These types of fields are
now beginning to be observed in fully convective pre-MS stars
\cite{gregory_2009}. Therefore, populations of stars with associated
discs should show significant differences in accretion signatures
across the CR transition region. Furthermore, as the X-ray emission
characteristics change across the CR transition we would expect this
to result in a significantly different interaction between star and
disc in terms of X-ray evaporation of the outer disc
\citep{ercolano_2009}. This may result in detectable differences in
mass loss from systems either side of the CR transition.

\section[]{ACKNOWLEDGMENTS}
NJM is supported by STFC grant ST/F003277/1. The author would like to
thank Eli Bressert, Rob King, Tim Naylor and Tim Harries for valuable
suggestions and discussion. Additionally, the author would like to
thank the referee for their suggestions resulting in a much improved
work.

\bibliographystyle{mn2e}
\bibliography{references}
\appendix

\label{lastpage}
\end{document}